\documentclass{mn2e}
\usepackage{psfig}
\usepackage{epsf}
\newcommand{\ltaraw}{$\; \buildrel < \over \sim \;$}
\newcommand{\lta}{\lower.5ex\hbox{\ltaraw}}
\newcommand{\gtaraw}{$\; \buildrel > \over \sim \;$}
\newcommand{\gta}{\lower.5ex\hbox{\gtaraw}}

\loadboldmathitalic 
\title [Microlensing of Fractal Sources]
{Gravitational microlensing of fractal sources} 
\author[Geraint F. Lewis]
{Geraint F. Lewis \\ 
Institute of Astronomy, School of Physics,
              University of Sydney, NSW 2006, Australia: {\tt gfl@physics.usyd.edu.au}
}
\date{\today}
\begin{document} 
\maketitle 
\begin{abstract}
Gravitational microlensing  has proven  to be a  powerful tool  in the
study of quasars, providing some of the strongest limits on the scales
of structure in the  central engine.  Typically sources are considered
to be smoothly  varying on some particular scale;  such simple sources
result in  recognisable time scales in microlensing  light curves from
which  the size  of the  source can  be determined.   Various emission
processes,  however,  result in  sources  with  a fractal  appearance,
possessing structure  on a range  of scales.  Here,  the gravitational
microlensing  of such  fractal  sources  at the  heart  of quasars  is
considered.  It is  shown that the resulting light  curves reflect the
fractal  nature of  the  sources, possessing  pronounced structure  at
various  scales,  markedly  different  to  the case  with  the  random
distribution of emission clouds that are typically considered.  Hence,
the  determination  of a  characteristic  scale  of  variability in  a
microlensing light  curve may not  necessarily reveal the size  of the
individual  emission  clouds,  the  key  value  that  is  required  to
determine the  physical state  of the emission  region, rather  it may
correspond to a particular  hierarchy in a fractal structure.  Current
X-ray satellites can detect  such fractal structure via the monitoring
of  gravitationally  lensed   quasars  during  a  microlensing  event,
providing a test of high energy emission processes in quasars.
\end{abstract}

\section{Introduction}\label{introduction}
Quasars represent some  of the most luminous objects  in the universe.
While  their spectra  reveal  clues to  the  various emission  regions
within the  active nucleus, their  small angular size  at cosmological
distances means that  we are not able to  directly image their central
regions.  New clues  have come from quasars which  have been magnified
by gravitational microlensing, and observations of microlensing events
have provided strong  constraints on the relative sizes  of the quasar
emission  regions (Rauch \&  Blandford 1991;  Lewis, Irwin,  Hewett \&
Foltz 1998; Agol, Jones \& Blaes 2000; Yonehara 2001).  Recent studies
have focused  upon the  influence of obscuring  material in  the broad
absorption  line region  (Lewis \&  Belle 1998)  and in  the extensive
broad  emission line  region (Wyithe  \& Loeb  2002) as  the continuum
source is microlensed. These studies  have found that the shadowing of
the emission/absorption  clouds produces pronounced  signatures on the
microlensing light  curve, and  the scale of  these signatures  can be
used to infer the size of the obscuring clouds.

\begin{figure*}
\centerline{ \psfig{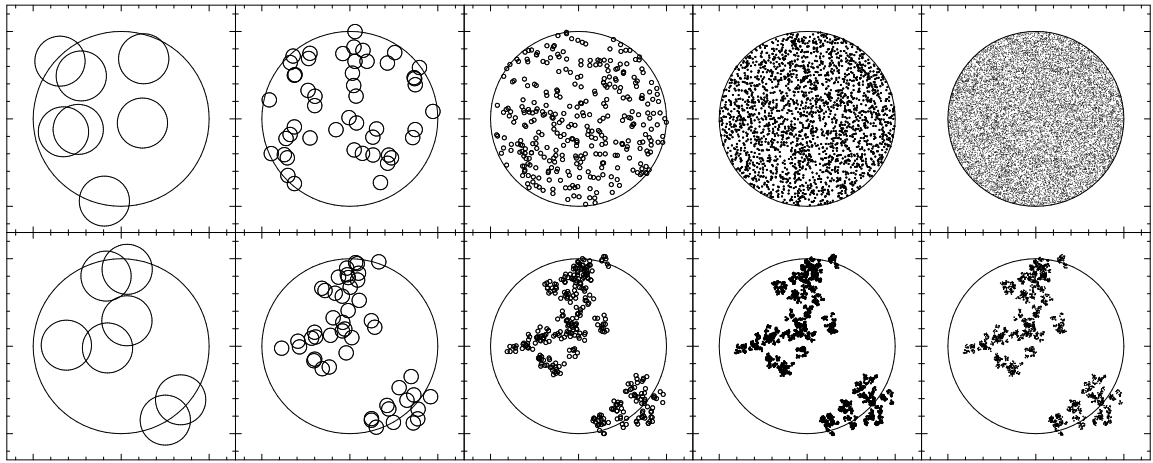}}
\caption[]{The  lower  row of  panels  presents  a  series of  fractal
hierarchies  for  the  source   structure  discussed  in  this  paper,
beginning  with level  1  on the  left-hand  side to  level  5 on  the
right-hand  side.    The  upper  series  of   panels  presents  random
distributions of the same number and  size of clouds seen in the lower
panels.   At the  lowest  hierarchical levels  (left-hand panels)  the
distributions  are similar  in the  random and  fractal cases,  but as
higher hierarchical levels are considered, the clumping in the fractal
distribution ensures it rapidly  deviates from the random distribution
of clouds.  The large circle in  each panel corresponds  to the lowest
(zeroth) fractal heirarchy, of radius $R_{max}$.
\label{fig1}}
\end{figure*}

Typically in microlensing studies, emission and absorption regions are
deemed to  have a  simple brightness profile  and scale,  and problems
focus upon determining these  scales from observations of fluctuations
induced by  gravitational microlensing.  In recent  years, however, it
has been realised  that many natural processes are  fractal in nature,
possessing similar structural properties on a variety of scales.  This
paper presents an investigation  of the influence of fractal structure
on the properties of microlensing light curves, especially with regard
to  the  determination of  the  scale  of  structure of  a  particular
emission region.  In  Section~\ref{fractal}, fractals in astronomy are
briefly  reviewed,  outlining the  physical  motivation  of the  model
adopted  in this  paper, while  in Section~\ref{lensing}  the approach
adopted to  study the gravitational microlensing of  fractal clouds is
presented, whereas the results are presented in Section~\ref{results}.
The conclusions are discussed in Section~\ref{conclusions}.

\section{Astronomical Fractals \& Fractal Sources}\label{fractal}
The fact that  a large number of natural  phenomenon appear to possess
fractal structure,  displaying self-similarity  on a range  of scales,
has been known  for sometime (e.g. Peitgen \&  Richer 1986; Peitgen \&
Saupe   1988).   Such   structure   has  been   observed  in   various
astrophysical  contexts,  from  the  surface  of  Mars  (Stepinski  et
al. 2002), the distribution of asteroids (Campo Bagatin et al.  2002),
and the influence of fractal  dust grains (Wright 1989; Fogel \& Leung
1998).  One of  the longest running debates has  involved the question
of  whether  large scale  structure  in  the  universe is  fractal  in
nature~\footnote{See  {\tt http://pil.phys.uniroma1.it/astro.html} for
a description of  the debate.}, as such a  conclusion would invalidate
the  central ideas of  relativistic cosmology  (e.g. Durrer  \& Labini
1998).

Many  astrophysical   studies  of  fractal   distributions  have  been
concerned with  self-similar structure  in gas clouds  (e.g. Elmegreen
1997; Stutzki  et al.  1998;  Elmegreen et al.  2001;  Elmegreen 2002;
Datta 2003) or in stellar distributions (Elmegreen \& Elmegreen 2001).
While these studies may  seem somewhat esoteric, with fractal analyses
providing useful  classification tools  (e.g.  Lekshmi et  al.  2003),
the existence  of fractal structure  influences the evolution  of both
gaseous  (Semelin \& Combes  2002) and  stellar (Goodwin  \& Whitworth
2004) components. Furthermore, Bottorff \& Ferland (2001) examined the
broad-line   region    of   quasars,   suggesting    that   transient,
turbulence-induced  overdensities  within  the emission  region  would
possess  fractal  structure,  complicating  the interpretation  of  the
physical properties of the region.

Fractal processes  have also been  associated with Sun,  including its
global   activity  (Salakhutdinova   1998),  radio   bursts  (Veronig,
Messerotti  \&   Hanslmeier  2000)  and   large-scale  magnetic  field
(Burlaga, Wang  \& Ness 2003). The small-scale  magnetic structure too
appears to  possess many fractal features (Abramenko  1999; Stenflo \&
Holzreuter 2003;  Jan{\ss}en, V\"{o}gler \& Kneer  2003).  In quasars,
the  X-ray emission  arises from  the most  central regions,  amidst a
complex  and dynamic  magnetic field  structure in  an  accretion disk
corona (e.g.  Merloni  \& Fabian 2001a), similar to  the activity seen
in the Solar  corona. Cascades of activity in  such regions may result
in fractal  like emission (Merloni  \& Fabian 2001b).  Given  that the
small X-ray emitting region at  the centres of quasars are amenable to
microlensing (Yonehara et al.  1998), this paper explores microlensing
signatures  of a quasar  X-ray emission  region with a  fractal surface
brightness distribution.

\begin{figure*}
\centerline{ \psfig{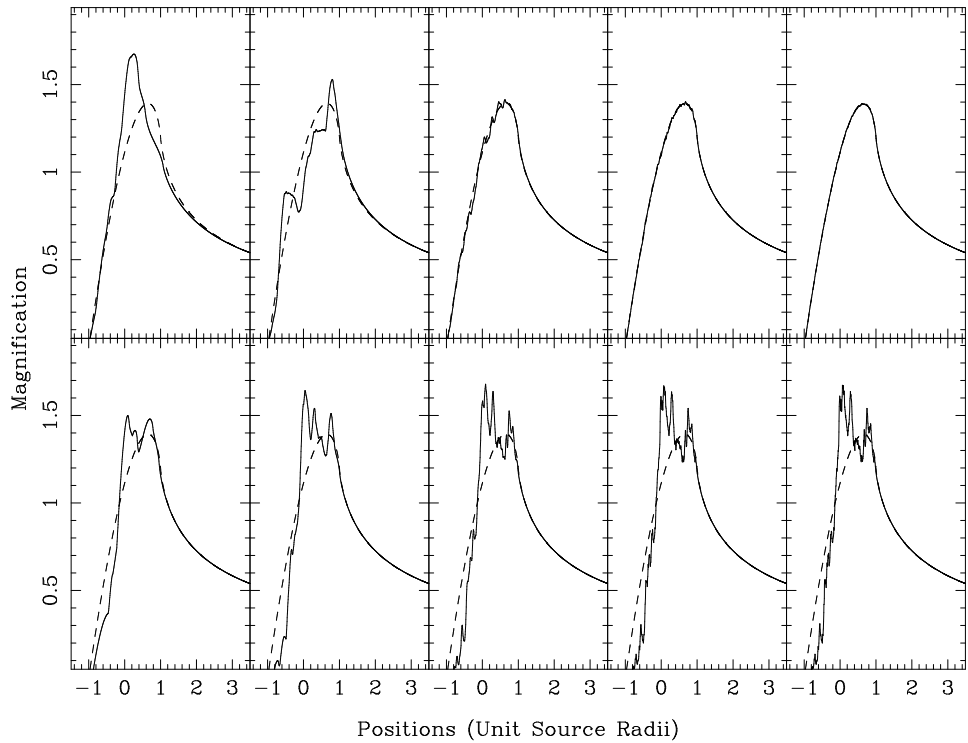}}
\caption[]{The result of sweeping the  caustic from left to right over
the cloud distributions presented in Figure~\ref{fig1}. In each panel,
the dashed line corresponds to the light curve for a uniform source of
unit radius.   In the upper  panels, which present the  combined light
curves of randomly distributed  sources, the total light curve rapidly
matches that  of the single  uniform source. The lower  panels present
the light  curves for the fractally distributed  sources. Clearly, the
behaviour of  this light curve is  different to the  random case.
\label{fig2}}
\end{figure*}

\section{Method}\label{method}

\subsection{Gravitational Microlensing}\label{lensing}
The fluctuations in brightness induced by a gravitational lensing mass
depend implicitly  on its  caustic structure. An  isolated, point-like
mass, such as a MaCHO  within the Galactic halo, produces a point-like
caustic.   This results  in  a simple,  smoothly  varying light  curve
(e.g. Alcock et al.  1993).  Multiple stars, however, can combine in a
very non-linear fashion, leading to an extended caustic structures and
more complex  light curves  (e.g.  Alcock et  al.  2000).   At optical
depths around unity, where many individual lenses combine to influence
the light from a distant  source, the resulting caustic structure, and
hence brightness fluctuations, are  quite complex (see Kayser, Refsdal
\&  Stabell 1986; Wambsganss  1992); this  situation is  applicable to
microlensed quasars such as  Q2237+0305 (Wambsganss, Paczynski \& Katz
1990). An examination of  such complex caustic structures reveals that
they are  dominated by fold  catastrophes which link the  higher order
catastrophes  (Witt 1990).   For the  purposes  of this  paper, it  is
assumed that the  source under consideration is small  compared to the
overall caustic structure, such that the source an be seen to be swept
by  single a  fold caustic,  an assumption  which hold  for  the X-ray
emitting  regions of  quasars  (Yonehara  et al.  1998).   It is  also
assumed  that the  caustic is  straight on  the scale  of  the overall
source structure,  such that higher order curvature  effects (Fluke \&
Webster 1999) can be neglected. With this, the magnification of a fold
caustic is simply given as;
\begin{equation}
\mu(x) = \sqrt{ \frac{ g }{x - x_c}} H(x-x_c) + \mu_o
\label{magnification}
\end{equation}
where $x-x_c$  represents the  perpendicular distance to  the caustic,
$H(x)$ is the  Heaviside step function and $g$  is the ``strength'' of
the caustic (Witt  1990). The $\mu_o$ term accounts  for the flux from
the other  images caused by the  caustic network and is  assumed to be
constant in the vicinity of the  caustic; this is taken to be zero for
this study  without any loss of  generality.  The clouds~\footnote{For
this paper, clouds  refers to regions of emission.   While the studies
of  Lewis  \&  Belle  (1998)  and Wyithe  \&  Loeb  (2002)  considered
absorbing clouds in front of a continuum source, the results presented
in  this paper  can be  scaled to  this case.}   are considered  to be
circular  and  of  uniform  surface brightness.   In  calculating  the
resulting  brightness fluctuation  as  a  source is  swept  by a  fold
caustic, the  analytic expressions of Schneider \&  Wagoner (1987) are
employed, setting their limb darkening parameter, $c_+=0$.

Chang (1984) demonstrated  that a source of radius  $R$ which is swept
by a fold caustic suffers a maximum magnification of;
\begin{equation} 
\mu_{max} = \sqrt{ \frac{ g }{R} } f
\label{max}
\end{equation}
where $f$ is a form factor,  which accounts for the specific shape and
surface brightness distribution of a source. For the uniform, circular
sources  under  consideration in  this  paper,  $f=1.39$. The  caustic
strength $g$ is  assumed to be unity throughout  this analysis. Detail
of  the  physical  scaling  of  this model  to  observed  microlensing
situations can be found in Lewis \& Belle (1998).

\subsection{Cloud Distributions}\label{clouds}
In  producing a  fractal  cloud distribution,  several parameters  are
required; the geometric factor ($L$) relates the scale of substructure
within a  structure, such  that a structure  of size $R$  will possess
substructures of size $R L^{-1}$. The multiplicity, $N$, is the number
of such substructures within  a structure; these parameters define the
fractal dimension  of the distribution  with $D = \log{N}  / \log{L}$.
The maximum  hierarchy, $H$, is  the number of scales  of substructure
considered.   With  this, the  radius  of  the smallest  substructures
(equivalent to individual clouds) is;
\begin{equation}
R_{cloud} = R_{max} L^{-H}
\label{fractaldist}
\end{equation}
where $R_{max}$ is  the radius of the largest  scale of the hierarchy,
corresponding to  the radius of the overall  region being microlensed.
Similarly, the scale of each hierarchy is;
\begin{equation}
R(h) = R_{max} L^{-h}, \; h=0,1\ldots H
\label{strucsize}
\end{equation}
where $h=H$  corresponds to  the individual clouds,  and $h=0$  is the
lowest level  of the fractal  hierarchy, corresponding to a  radius of
$R_{max}$.

Hence, in generating the  fractal distribution, $N$ subregions of size
$R=R_{max}  L^{-1}$  are  randomly  scattered into  the  zeroth  level
hierarchy, within a radius of $R_{max}$.  Within these subregions, $N$
further  subregions of  size  $R=R_{max} L^{-2}$  are scattered.   The
procedure is  continued until the subregions  represent the individual
clouds  themselves  [see Bottorff  \&  Ferland  (2001)  for a  further
description of this procedure].

To date, there are no detailed models to predict precise fractal X-ray
surface  brightness  distributions   within  the  central  regions  of
quasars.  To  this end, the  fractal distributions presented  here are
illustrative only, not driven  by any particular physical constraints.
A fractal  model was  chosen with $L=3.5$  and a fractal  dimension of
$D=1.55$, corresponding to $7$  structures per level. The lower panels
of Figure~\ref{fig1}  presents the spatial distributions  of the could
distributed with these fractal parameters; from the left to right, the
panels present the first to  the fifth hierarchy of fractal structure;
the  large circle  in all  panels corresponds  to the  lowest (zeroth)
level of  the fractal  hierarchy and has  a radius of  $R_{max}$.  The
lower left-most panel presents the  case with $h=1$, the first fractal
hierarchy.  Here there  are  seven source  regions randomly  scattered
within $R_{max}$,  each with a  radius of $R_{cloud} =  R_{max} L^{-1}
\sim 0.29 R_{max}$.  In the panel immediately to the right, the second
fractal  hierarchy, $h=2$,  is considered,  and each  of  the emission
clouds  in the  $h=1$  level has  been  broken up  into seven  smaller
clouds, each  with a radius of  $R_{cloud} = R_{max}  L^{-2} \sim 0.08
R_{max}$, with  each cloud being randomly scattered  within the circle
of the  $h=1$ cloud.  This process to  $h=5$ in the  right-most panel.
This  right-most  panel consists  of  $7^5=16807$ individual  emission
clouds, each with  a radius of $R_{cloud} =  R_{max} L^{-5} \sim 0.002
R_{max}$.  The upper panels  consider a similar procedure, but instead
of scattering  clouds in each hierarchy within  the boundaries defined
by the  hierarchy below it,  the clouds are scattered  randomly within
$R_{max}$,   providing  an   overall  random   distribution.    It  is
immediately   apparent   that   the   fractal   distributions   differ
significantly from the random distributions and possess structure on a
number of scales (as expected for fractal distributions).

\section{Results}\label{results}

\subsection{Microlensing Light Curves}\label{results2}
Figure~\ref{fig2}  presents  the  microlensing  light curves  for  the
sources at  each hierarchy  presented in Figure~\ref{fig1};  note that
for the  simulations presented in this  paper, it is  assumed that the
source  region remains  fixed  and unvarying  as  it is  swept by  the
microlensing  caustic; such variability  would further  complicate the
resulting  light  curve, confusing  the  microlensing signature.   The
dashed line in  each panel corresponds microlensing light  curve for a
source of  unit radius, corresponding  to $R_{max}$ in this  case, and
the surface brightness  of each source hierarchy has  been adjusted so
that  it matches this  light curve  after the  caustic has  passed the
emission region. In the left-most panel, the source region consists of
a  small number  of relatively  large  sources, a  situation which  is
reflected in  the light  curve, which exhibits  substantial variations
about the unit radius source light curve.  Moving towards the right of
the top panels of Figure~\ref{fig2},  as the source size decreases and
their  number increases,  the size  of  the deviations  from the  unit
source  decreases, becoming imperceptible  in the  final frame  of the
panel,  representing  the  resultant  light curve  of  16807  randomly
distributed sources.

Considering  the lower  panels of  Figure~\ref{fig2} reveals  that, as
expected, the lowest fractal  hierarchy produces a similar light curve
to the random  distribution of sources, a situation  which is apparent
in  the second  hierarchy.  In  moving to  higher  fractal hierarchies
(towards the  right), however, it is  seen that there  is a pronounced
difference between the light curves from the fractal distributions and
the random distributions, with the light curves of the fractal sources
not approaching that  of the unit circular source.   Rather, as higher
hierarchies are considered, it is  seen that structures that appear in
lower hierarchies persists.

This  behaviour is  straight  forward to  understand;  for the  random
distributions, additional sources added  at each hierarchy act to {\it
fill  the gaps}  in the  light curve,  eventually smoothing  the light
curve out  to the uniform source  (as seen in the  right-most panel of
Figure~\ref{fig2}).   For  the fractal  source,  however, the  smaller
sources  in the  higher  hierarchies  are constrained  to  lie in  the
regions  of  the lower  hierarchies.  This  ensures  that each  higher
fractal level  effectively maintains  the larger scale  variability of
the preceding hierarchy, adding smaller scale variability to it. 

Note  that  in  this  example  the fractal  hierarchy  has  only  been
considered down to fifth level.  In reality, the fractal hierarchy can
be substantially deeper [c.f. Bottorff \& Ferland (2001) who find that
the broad line region of quasars is best represented by a fractal with
eleven hierarchies, corresponding  to a total $\sim10^{10}$ individual
emission clouds]. If this is  the case, then the light curve presented
for the random distributions would appear quite smooth and would do so
until explored on the scale of the very high hierarchies.  The fractal
distribution would,  however, continue  to display variability  on all
the scales of the fractal distribution.

\subsection{Implications}\label{implications}
A  number of studies  have attempted  to determine  the scale  size of
continuum  emitting  region  in   quasars  using  the  time  scale  of
gravitational  microlensing  variability  (Wyithe, Webster  \&  Turner
2000; Yonehara et al. 1999; Kochanek 2004), including specifically the
X-ray  emitting  region  (Yonehara  et  al.   1998;  Popovi{\'  c}  et
al. 2003). All these studies, however, assume a simple source profile.

Would a  fractal source result  in any observational  consequences? To
examine  this  it   is  important  to  consider  the   time  scale  of
microlensing  variability.   For the  quadruple  lens Q2237+0305,  the
caustic crossing time scale is
\begin{equation}
t \sim \left( \frac{ r_{src} }{4\times10^{13}cm}\right)
\left( \frac{ v_t }{600 km/s}\right)^{-1} \ \ \ days
\label{vels}
\end{equation}
where  $r_{src}$  is  the  radius  of  the source  and  $v_t$  is  the
transverse  velocity  of  the  lensing  galaxy (Kayser  et  al.  1998;
Yonehara  etc  al.  1998).   Confined  to the  inner  regions  of  the
accretion  disk,  the  X-ray   emitting  region  is  estimated  to  be
$\sim10-20$  Schwarzschild radii  in extent  (e.g.  Merloni  \& Fabian
2001),  corresponding to  $\sim3\times10^{14}$cm  for a  $10^8M_\odot$
black  hole.  Clearly,  given  Equation~\ref{vels}, the  light  curves
presented in  Figure~\ref{fig2} are $\sim1$ month in  duration.  It is
important  to note,  however, that  all  the light  curves are  smooth
beyond unity along  the spatial axis, corresponding to  times when the
sharp boundary  of the  fold caustic has  swept completely  across the
entire source  region.  Hence, strong  variability due to  the fractal
structure of the source can  only be seen when this high magnification
region lies across  the source, confining the strong  variability to a
period of a  couple of weeks. To obtain  a signal-to-noise of $\sim10$
in the  faintest images of  Q2237+0305, Chandra need to  integrate for
$\sim15$ks  (Dai  et al.  2003),  implying  a  temporal resolution  of
$\sim4-5$ hours and, via  Equation~\ref{vels}, a spatial resolution of
$\sim7\times10^{12}$cm;  this corresponds  to  the third  hierarchical
level  ($h=3$)  for  the   fractal  distribution  considered  in  this
paper.  Hence, with this  temporal sampling  and 10\%  photometry, the
fractal nature of the source  would be imprinted on the observed light
curve.

Given the self-similar structure in the source, and hence in the light
curve, any  observing period shorter  than the total  caustic crossing
displayed  in  Figure~\ref{fig2}  may  identify  variability  that  is
interpreted as revealing the scale  size of the X-ray emission region,
whereas  it truly  corresponds to  a particular  level in  the fractal
hierarchy; this will be especially true if the data are noisy, masking
any variations  introduced due to the substructures  in higher fractal
hierarchies. In  searching for the full  fractal signature, therefore,
it is imperative to following the entire sweeping of the source region
during a microlensing event.  As the optical/UV emission region should
be  substantially   larger  than   the  X-ray  emitting   region,  any
microlensing in the optical/UV  should possess a proportionally larger
time scale.   Hence, monitoring at  these longer wavelengths  could be
used to to identify microlensing events, and providing a trigger for 
monitoring in the X-ray. 

Of  course, it  is important  to  note that  the fractal  distribution
discussed in  this paper does not  represent a specific  model for the
X-ray surface brightness in  quasars.  While the particular imprint of
a  fractal source  on  a  microlensing light  curve  depends upon  the
details of the  distribution, it is expected that  any resulting light
curve will possess  self-similar structure on a range  of scales. This
can be  used to determine the  physical validity of  models for quasar
X-ray emission (e.g. Merloni \& Fabian 2001b).

\section{Conclusions}\label{conclusions}
Studies of the gravitational  microlensing of quasars typically assume
that source are smooth, possessing  a typical scale-length that can be
determined  from observations of  microlensing variability.   Over the
last decade, it  has become clear that a  number of physical processes
can result  in structures with  fractal properties. This  has included
high energy emission from the solar corona and, in an analogous model,
to X-ray emission from the heart of quasars.

This paper  has considered the  influence of fractal structure  in the
X-ray emitting region of quasars as  it is swept by a caustic during a
microlensing event.  It  was found that, for a  random distribution of
clouds, as  the size  of the clouds  were decreased, and  their number
increased, the total  light curve rapidly approached that  of a larger
uniform  source.  Sources  distributed  fractally, however,  displayed
quite different  properties; as the  source size is decreased  and the
number  of clouds increased,  they are  not distributed  randomly, but
within  fractal  hierarchies.  This  clustering  of  sources is  quite
apparent in the resultant light curve, with higher fractal hierarchies
adding variability  substructure to the lower  hierarchies. Unlike the
randomly distributed  sources, the light curve does  not resemble that
of a  larger, uniform source,  but (like the source  itself) possesses
substructure on  a range  of scales.  An  examination of  the relevant
time  scales reveals  that monitoring  of  gravitationally microlensed
quasars  with X-ray satellites  would allow  the determination  of the
lower  levels of  fractal structure,  providing further  tests  to the
underlying emission mechanisms.

The  small,  X-ray  emitting  regions  of quasars  are  not  the  only
structures  in active  galaxies that  possess fractal  structure, with
Bottorff and  Ferland (2001) suggesting that the  extensive broad line
emitting region should also be  fractal in nature. Microlensing of the
broad line  region was  examined by Nemiroff  (1988) and  Schneider \&
Wambsganss  (1990) who  found  that due  to  the overall  size of  the
emission region ($\sim$parsec scale) the  flux in the emission line is
essentially unaltered due to the action of gravitational microlensing,
although microlensing could result in variability of the emission line
profiles.  Reverberation mapping experiments have suggested an overall
smaller size  for the broad  line region ($\sim0.1$  parsecs), further
indicating that the region  should be amenable to strong gravitational
microlensing   (Abajas   et  al.    2004;   Lewis   \&  Ibata   2004).
Observational  evidence  for this  has  recently  been reported,  with
Richards et al.  (2004) identifying strong line profile variability in
the gravitationally lensed quasar SDSS J1004+4112 which they attribute
to  the action  of  microlensing.   If, as  suggested  by Bottorff  \&
Ferland  (2001),  the broad  line  region  of  quasars also  possesses
fractal  structure, it  too may  be  susceptible to  the influence  of
gravitational microlensing.  Given its size, however, this region will
cover the complex caustic network seen in microlensing, and the simple
analysis presented in this  paper will not apply.  Requiring numerical
simulation  of high  optical  depth microlensing  (Kayser, Refsdal  \&
Stabell 1986; Wambsganss  1992), this will form the  basis for further
study.

\section*{Acknowledgements}
GFL  thanks   Zdenka  Kuncic  for  enlightening   discussions  on  the
generation of X-rays in quasars  and for suggestions that improved the
clarity  of this  paper. The  anonymous  referee is  also thanked  for
comments which improved the paper.

\newcommand{\aap}{A\&A}
\newcommand{\apj}{ApJ}
\newcommand{\apjl}{ApJ}
\newcommand{\aj}{AJ}
\newcommand{\mnras}{MNRAS}
\newcommand{\apss}{Ap\&SS}
\newcommand{\nat}{Nature}


\begin{thebibliography}{DUM}
\bibitem[\protect\citeauthoryear{Abajas et al.}{2002}]{2002ApJ...576..640A} 
Abajas C., Mediavilla E., Mu{\~ n}oz J.~A., Popovi{\' c} L.~{\v C}., Oscoz 
A., 2002, ApJ, 576, 640 

\bibitem[\protect\citeauthoryear{Abramenko}{1999}]{1999ARep...43..622A} 
Abramenko V.~I., 1999, ARep, 43, 622 

\bibitem[Alcock et al.~1993]{1993Natur.365..621A} 
Alcock C., Akerloff C.~W., Allsman R.~A., et al., 
1993, \nat,  365, 621

\bibitem[Alcock et al.~2000]{2000ApJ...541..270A} 
Alcock C., Allsman R.~A., Alves D., et al., 
2000, \apj,  541, 270

\bibitem[Agol et al.~2000]{2000ApJ...545..657A} 
Agol E., Jones B., Blaes O., 
2000, \apj,  545, 657

\bibitem[\protect\citeauthoryear{Burlaga, Wang, \& 
Ness}{2003}]{2003GeoRL..30j..50B} Burlaga L.~F., Wang C., Ness N.~F., 2003, 
GeoRL, 30, 50 

\bibitem[Bottorff \& Ferland 2000]{2000MNRAS.316..103B} 
Bottorff M.~C., Ferland G.~J., 
2000, \mnras,  316, 103

\bibitem[Bottorff \& Ferland 2001]{2001ApJ...549..118B} 
Bottorff M., Ferland G., 
2001, \apj,  549, 118

\bibitem[\protect\citeauthoryear{Campo Bagatin, Mart{\'{\i}}nez, \& 
Paredes}{2002}]{2002Icar..157..549C} Campo Bagatin A., Mart{\'{\i}}nez 
V.~J., Paredes S., 2002, Icar, 157, 549 

\bibitem[Chang 1984]{1984A&A...130..157C} 
Chang K., 
1984, \aap,  130, 157

\bibitem[\protect\citeauthoryear{Dai et al.}{2003}]{2003ApJ...589..100D} 
Dai X., Chartas G., Agol E., Bautz M.~W., Garmire G.~P., 2003, ApJ, 589, 
100 

\bibitem[\protect\citeauthoryear{Datta}{2003}]{2003A&A...401..193D} Datta 
S., 2003, A\&A, 401, 193 

\bibitem[\protect\citeauthoryear{Durrer \& 
Labini}{1998}]{1998A&A...339L..85D} Durrer R., Labini F.~S., 1998, A\&A, 
339, L85 

\bibitem[Elmegreen 1997]{1997ApJ...477..196E} 
Elmegreen B.~G., 
1997, \apj,  477, 196

\bibitem[\protect\citeauthoryear{Elmegreen}{2002}]{2002ApJ...564..773E} 
Elmegreen B.~G., 2002, ApJ, 564, 773 

\bibitem[\protect\citeauthoryear{Elmegreen \& 
Elmegreen}{2001}]{2001AJ....121.1507E} Elmegreen B.~G., Elmegreen D.~M., 
2001, AJ, 121, 1507 

\bibitem[\protect\citeauthoryear{Elmegreen, Kim, \& 
Staveley-Smith}{2001}]{2001ApJ...548..749E} Elmegreen B.~G., Kim S., 
Staveley-Smith L., 2001, ApJ, 548, 749 

\bibitem[Fluke \& Webster 1999]{1999MNRAS.302...68F} 
Fluke C.~J., Webster R.~L., 
1999, \mnras,  302, 68

\bibitem[\protect\citeauthoryear{Fogel \& 
Leung}{1998}]{1998ApJ...501..175F} Fogel M.~E., Leung C.~M., 1998, ApJ, 
501, 175 

\bibitem[\protect\citeauthoryear{Goodwin \& 
Whitworth}{2004}]{2004A&A...413..929G} Goodwin S.~P., Whitworth A.~P., 
2004, A\&A, 413, 929 

\bibitem[\protect\citeauthoryear{Jan{\ss}en, V{\" o}gler, \& 
Kneer}{2003}]{2003A&A...409.1127J} Jan{\ss}en K., V{\" o}gler A., Kneer F., 
2003, A\&A, 409, 1127 

\bibitem[Kayser et al.~1986]{1986A&A...166...36K} 
Kayser R., Refsdal S., Stabell R., 
1986, \aap,  166, 36

\bibitem[\protect\citeauthoryear{Kochanek}{2004}]{2004ApJ...605...58K} 
Kochanek C.~S., 2004, ApJ, 605, 58 

\bibitem[\protect\citeauthoryear{Lekshmi, Revathy, \& Prabhakaran 
Nayar}{2003}]{2003A&A...405.1163L} Lekshmi S., Revathy K., Prabhakaran 
Nayar S.~R., 2003, A\&A, 405, 1163 

\bibitem[Lewis \& Belle 1998]{1998MNRAS.297...69L} 
Lewis G.~F., Belle K.~E., 
1998, \mnras,  297, 69

\bibitem[\protect\citeauthoryear{Lewis \& 
Ibata}{2004}]{2004MNRAS.348...24L} Lewis G.~F., Ibata R.~A., 2004, MNRAS, 
348, 24 

\bibitem[Lewis et al.~1998]{1998MNRAS.295..573L} 
Lewis G.~F., Irwin M.~J., Hewett P.~C., Foltz C.~B., 
1998, \mnras,  295, 573

\bibitem[\protect\citeauthoryear{Merloni \& 
Fabian}{2001}]{2001MNRAS.321..549M} Merloni A., Fabian A.~C., 2001a, MNRAS, 
321, 549 

\bibitem[\protect\citeauthoryear{Merloni \& 
Fabian}{2001}]{2001MNRAS.328..958M} Merloni A., Fabian A.~C., 2001b, MNRAS, 
328, 958 

\bibitem[Nemiroff 1988]{1988ApJ...335..593N} 
Nemiroff R.~J., 
1988, \apj,  335, 593

\bibitem[sd]{wwe}
Peitgen, H.-O., Saupe, D., 1988, {\it The Science of Fractal Images}, 
Springer-Verlag (Berlin)

\bibitem[sd]{awe}
Peitgen, H.-O., Richter, P.~H., 1986, {\it The Beauty of Fractals}, 
Springer-Verlag (Berlin)

\bibitem[\protect\citeauthoryear{Popovi{\' c} et 
al.}{2003}]{2003A&A...398..975P} Popovi{\' c} L.~{\v C}., Mediavilla E.~G., 
Jovanovi{\' c} P., Mu{\~ n}oz J.~A., 2003, A\&A, 398, 975 

\bibitem[Rauch \& Blandford 1991]{1991ApJ...381L..39R} 
Rauch K.~P., Blandford R.~D., 
1991, \apj,  381, L39

\bibitem[wewew]{dfdfdf}
Richards G. T. et al.,
2004, \apj, 610, 679

\bibitem[\protect\citeauthoryear{Salakhutdinova}{1998}]{1998SoPh..181..221S} 
Salakhutdinova I.~I., 1998, SoPh, 181, 221 

\bibitem[Schneider \& Wagoner 1987]{1987ApJ...314..154S} 
Schneider P., Wagoner R.~V., 
1987, \apj,  314, 154

\bibitem[Schneider \& Wambsganss 1990]{1990A&A...237...42S} 
Schneider P., Wambsganss J., 
1990, \aap,  237, 42

\bibitem[\protect\citeauthoryear{Semelin \& 
Combes}{2002}]{2002A&A...387...98S} Semelin B., Combes F., 2002, A\&A, 387, 
98 

\bibitem[\protect\citeauthoryear{Stenflo \& 
Holzreuter}{2003}]{2003AN....324..397S} Stenflo J.~O., Holzreuter R., 2003, 
AN, 324, 397 

\bibitem[\protect\citeauthoryear{Stepinski et 
al.}{2002}]{2002LPI....33.1347S} Stepinski T.~F., Marinova M.~M., McGovern 
P.~J., Clifford S.~M., 2002, LPI, 33, 1347 

\bibitem[\protect\citeauthoryear{Stutzki et 
al.}{1998}]{1998A&A...336..697S} Stutzki J., Bensch F., Heithausen A., 
Ossenkopf V., Zielinsky M., 1998, A\&A, 336, 697 

\bibitem[\protect\citeauthoryear{Veronig, Messerotti, \& 
Hanslmeier}{2000}]{2000A&A...357..337V} Veronig A., Messerotti M., 
Hanslmeier A., 2000, A\&A, 357, 337 

\bibitem[Wambsganss 1992]{1992ApJ...386...19W} 
Wambsganss J., 
1992, \apj,  386, 19

\bibitem[Wambsganss et al.~1990]{1990ApJ...352..407W} 
Wambsganss J., Paczynski B., Katz N., 
1990, \apj,  352, 407

\bibitem[Witt 1990]{1990A&A...236..311W} 
Witt H.~J., 
1990, \aap,  236, 311

\bibitem[Wright 1989]{1989ApJ...346L..89W} 
Wright E.~L., 
1989, \apj,  346, L89

\bibitem[aa]{2002ApJ...577..615W} 
Wyithe J.~S.~B., Loeb A., 2002, \apj, 577, 615 

\bibitem[\protect\citeauthoryear{Wyithe, Webster, \& 
Turner}{2000}]{2000MNRAS.318..762W} Wyithe J.~S.~B., Webster R.~L., Turner 
E.~L., 2000, MNRAS, 318, 762 


\bibitem[Yonehara 2001]{2001ApJ...548L.127Y} 
Yonehara A., 2001, \apj,  548, L127

\bibitem[\protect\citeauthoryear{Yonehara et 
al.}{1999}]{1999A&A...343...41Y} Yonehara A., Mineshige S., Fukue J., 
Umemura M., Turner E.~L., 1999, A\&A, 343, 41 

\bibitem[\protect\citeauthoryear{Yonehara et 
al.}{1998}]{1998ApJ...501L..41Y} Yonehara A., Mineshige S., Manmoto T., 
Fukue J., Umemura M., Turner E.~L., 1998, ApJ, 501, L41 
\end{thebibliography}
\end{document}